\documentclass{evn2004}
\usepackage{txfonts}
\begin{document}
   \title{East Asian VLBI Activities}

   \author{M. Inoue\inst{1}
          }

   \institute{National Astronomical Observatory of Japan, Japan
             }

   \abstract{
The VLBI activities in the East Asia region are planning to coordinate among VLBI groups and institutes in China, Japan, and Korea.  In these countries, some new projects are going to start or discussed, and some types of collaboration have been already done individually.  These new projects are briefly introduced.  Under these circumstances, an organization has been discussed for the coordinated efforts to produce fruitful results more effectively.
   }

   \maketitle
%

\section{Introduction}

In the East Asia region of China, Japan, and Korea, there have been many activities in VLBI.  These activities are widely spread from system designing and developments of VLBI instruments, to observations in various fields of VLBI science and applications.  In China, two 25-m telescopes have been deeply involved in several VLBI networks, e.g., EVN and Asia-Pacific Space Geodynamics (APSG) by which a global deformation of Asian continent has been observed.  They have been building correlators and Hydrogen maser clocks.  In Japan, an innovative concept of correlator, FX, was proposed and built one for the first space VLBI project VSOP (VLBI Space Observatory Program).  By the Key Stone Project (KSP), a sudden movement was detected within a small area around Tokyo in July 2000.  KSP is the first optical fiber-linked VLBI network with a real time correlator.  

Recently, new VLBI networks have been built in Japan and Korea:  They are the VLBI Exploration of Radio Astrometry (VERA) and the Korean VLBI Network (KVN).  Their systems are oriented at higher frequencies and have several common features.  Encouraged to these new networks, studies to collaborate in VLBI have been discussed between China, Japan, and Korea.  To make the collaboration more efficiently, we are going to form a committee by these countries.

Here in this presentation, I will introduce activities in these countries, mainly from astronomical point of view, and give a short report on the status of the coordination.


\section{Activities in each country}

A brief introduction of VLBI activities is shown here for each country in alphabetical order.  Although there have been some interaction and collaboration between these countries, observations and system developments were not well coordinated, except activities in APSG.  

\subsection{In China}

Geodetic VLBI observations have been done by Shanghai Astronomical Observatory (SHAO) with the 25-m radio telescope, located near SHAO where the east coast of China.  Another 25-m radio telescope with the same design of that in SHAO was built and has been operated in Urumqi, located almost the North Western boundary of China, to form a 3000-km baseline.  The telescopes were originally designed for geodetic measurements, and SHAO has been performing APSG observations as the central office of APSG.  APSG observations show a global movement of the continent, presumably it being pushed up by Indian area.  

The telescopes have been upgraded to have high capability at 22 GHz, and participated in EVN observations to have an Eastern extension of the baselines.  SHAO has been developing VLBI instruments and a Hydrogen maser clock system.

The mm group in the Purple Mountain Observatory (PMO) is managing to make mm VLBI using the 14-m telescope in Delingha.  As each country has a mm telescope, it is possible to perform a mm VLBI network.

Recently, the space agency has a project to research the Moon using a VLBI technique, and then a VLBI station is under construction.

\subsection{In Japan}

The Communications Research Laboratories (CRL) has developed many models of VLBI terminal and correlators, and played an important role for defining the VLBI Standard Interface (VSI) specification.  The CRL also has been deeply involved in the International VLBI Survey for Geodesy and Astrometry (IVS) as a Technology Developement Center, Network Stations, Correlators, etc.  Among the activities in CRL, it should be noted that they built a four-station VLBI system to monitor a crustal deformation around Tokyo area, which was expected to see a precursor of a large earthquake in the metropolitan area with about 100-km baselines.  This KSP was operated for five years, and in later years the stations were connected by optical fiber to get real time fringes, which were essential to monitor any precursor of destructive earthquakes within a couple of days.

In recent days the CRL has been developing an IP VLBI system and a software correlation system using PC, which may change the situation and cost for the VLBI observing situation in near future.

The first space VLBI satellite HALCA was launched by the Institute of Space and Astronautical Science (ISAS) in close collaboration with the Nobeyama Radio Observatory (NRO) of NAOJ.  This project, named VSOP, has been performed by the world-wide collaboration in the VLBI community.  NAOJ is now running VERA, which consists of four 20-m radio telescopes in the Southern area of Japan to measure the annual parallax and proper motion of Galactic maser sources with an accuracy of 10 micro arcsec level using a phase referencing technique.  Some universities and institutes got radio telescopes to form a VLBI network with VERA.  With all of available telescopes in Japan, KVN, and those in China, we will have a unique VLBI network in the East Asia region.

The next space VLBI project VEOP-2 has been planned in Japan, and the East Asia VLBI network will be a good counterpart on the ground.  

A new application has been developed to navigate a space craft in deep space, which can be applied to measure a lunar orbiter to derive the potential of the gravity of the Moon.

\subsection{In Korea}

The KVN system is now under construction, and from this year they started to build a correlator for the system.  As the KVN and VERA systems have a lot of common features, it is natural to operate with a combined mode, together with all available telescopes in these countries.  For this, we need a large correlator to process such a multi station VLBI network.

Using the Taeduk 14-m mm radio telescope of Taeduk Radio Astronomy Observatory (TRAO), mm-VLBI observations has made at 86 GHz with the 45-m telescope of NRO.  Some SiO maser sources were observed with almost the East-West baseline of 1000 km.  Further 2000 km West of Taeduk, China has also the 14-m mm telescope in Delingha, and we are planing to make VLBI observations with these three mm telescopes.

\section{Coordination of these activities}

In recent years it becomes obvious that a close collaboration between China, Japan, and Korea could produce fruitful results.  The issues for the collaboration are extended in various things: coordinated joint VLBI observations, sharing of system developments, exchange of personnel and students, etc.  In particular, a data correlation center is needed if such coordinated observations are well performed.  To manage and promote these issues, we are discussing to have a standing committee.  We will have the first committee meeting in October this year.

To the South of the East Asia region, there are radio telescopes in Australia, and it would be a powerful network to combine both together, giving good East-West and North-South baselines.

\section{Conclusions}

Everything has a beginning and an end.  This is the beginning of the coordination among VLBI groups in these three countries.

\begin{acknowledgements}
The contents of this presentation are provided by many VLBI groups and institutes in China, Japan, and Korea.
\end{acknowledgements}







\end{document}